\documentclass[prd,aps,reprint,onecolumn,superscriptaddress,tightenlines,nofootinbib,eqsecnum,preprintnumbers,longbibliography,12pt]{revtex4-2}
\usepackage{epsfig,amsfonts,mathrsfs,amsmath,amssymb,graphicx,color,slashed,multirow}
\usepackage{amsmath,latexsym,amssymb,graphicx,slashed,hyperref,color,enumerate,url,etoolbox,cancel,gensymb,physics}
\hypersetup{colorlinks,citecolor= nicegreen,linkcolor= nicered}
\usepackage[usenames,dvipsnames]{xcolor}
\definecolor{nicered}{rgb}{0.7,0.1,0.1}
\definecolor{nicegreen}{rgb}{0.1,0.5,0.1}
\definecolor{niceblue}{rgb}{0.1,0.2,0.6}

\usepackage{bbm}
\usepackage{subfig}
\usepackage{physics}
\usepackage{dsfont}
\usepackage{cancel}
\usepackage[normalem]{ulem}

\begin{document}

\def\Carleton{Ottawa-Carleton Institute for Physics, Carleton University, Ottawa, ON K1S 5B6, Canada}
\def\UCSD{Department of Physics, University of California, San Diego, CA 92093, USA}

\title{On s-confining SUSY-QCD with Anomaly Mediation}








\author{Carlos Henrique de Lima}
\email{carloshenriquedelima@cmail.carleton.ca}
\affiliation{\Carleton}
\author{Daniel Stolarski}
\email{stolar@physics.carleton.ca}
\affiliation{\Carleton}
\affiliation{\UCSD}

\date{\today}

\begin{abstract}
In this work, we present a comprehensive study of the phase diagram of supersymmetric QCD with $N_{f}=N_{c}+1$ flavors perturbed by Anomaly Mediated Supersymmetry Breaking (AMSB). We extend the previous analyses on s-confining ASQCD theories in three different directions. We show that the existence of the QCD-like vacuum is independent of the size of the SUSY breaking parameter. We further expand the analysis of these models by including two and three-loop contributions to investigate the robustness and limitations of the results. Finally, we include the leading effect of higher-order Kähler terms to investigate the stability of the phase diagram as we approach the confining energy scale. The analysis with higher order Kähler terms is also extended for $N_{c}=2$ for which AMSB alone gives inconclusive results.
\end{abstract}


\maketitle
\section{Introduction}

Understanding the vacuum structure of strongly coupled gauge theories has been a longstanding goal in theoretical physics. One case of interest is theories like Quantum Chromodynamics (QCD) in the Standard Model. In particular, theories with an $SU(N_c)$ gauge group and $N_f$ vectorlike flavors of quarks in the fundamental representation. If $N_f \geq 2$, there is a global $SU(N_f)_L\times SU(N_f)_R \times U(1)_B$ symmetry in the absence of mass terms for the quarks. In the case where $N_f$ is sufficiently small that the theory is asymptotically free and not in the conformal window, then perturbation theory cannot be used to describe the low energy behavior of the theory. 

For actual QCD, one can use the hadron spectrum to deduce that there is spontaneous chiral symmetry breaking ($\chi$SB): $SU(N_f)_L \times SU(N_f)_R \rightarrow SU(N_f)_V$~\cite{Weinberg:1996kr}. It is widely believed that in general these theories will confine and spontaneously break chiral symmetry. There are a variety of arguments supporting this idea including those based on 't Hooft anomaly matching~\cite{tHooft:1979rat,Frishman:1980dq,Coleman:1982yg}, large $N_c$ arguments~\cite{Coleman:1980mx}, and the persistent mass condition~\cite{Preskill:1981sr,Schwimmer:1981yy,Farrar:1980sn,Takeshita:1981sx,Kaul:1981fd,Cohen:1981iz,Bars:1981nh,Ciambriello:2022wmh}. As yet, however, conclusive proof of $\chi$SB remains elusive. 

One can also study supersymmetric (SUSY) versions of QCD (SQCD) where dualities elucidated by Seiberg~\cite{Seiberg:1994bz,Seiberg:1994pq} allow one to obtain powerful exact results. One particular subclass of theories that we focus on in this work is those with $N_f=N_c+1$ that exhibit s-confinement. At low energy, this theory can be described by gauge invariant composite operators and a dynamical superpotential. This is analogous to how pions can describe QCD at low energy. In the absence of SUSY breaking, $N_f=N_c+1$ SQCD has a rich moduli space, and, at the origin of moduli space, chiral symmetry is \textit{not} broken in stark contrast to the non-supersymmetric lore and data. 

In order to attempt to use the SQCD to study QCD, one can introduce SUSY breaking~\cite{Aharony:1995zh,Cheng:1998xg,Arkani-Hamed:1998dti,Luty:1999qc,Abel:2011wv}. In the UV theory, when soft squark and gaugino masses are taken to infinity one recovers QCD. Unfortunately one needs the SUSY breaking to be in some sense small to be able to use the powerful supersymmetric results. In particular, if one specifies the squark and gaugino masses in the UV, there is in general no quantitative way to map those parameters onto SUSY breaking parameters in the low energy effective theory. This problem can be solved by using anomaly-mediated SUSY breaking (AMSB)~\cite{Randall:1998uk,Giudice:1998xp,Pomarol:1999ie,Jung:2009dg}. In AMSB, all the soft SUSY breaking terms depend on only a single parameter, the gravitino mass $m_{3/2}$, and the SUSY breaking is UV insensitive. This means that, for the case of s-confining SQCD, if the SUSY breaking is small compared to the dynamical scale, then one can describe SUSY breaking entirely in terms of gauge invariant composites in the low energy theory. 

Coupling AMSB to SQCD, a procedure called ASQCD, has been used~\cite{Murayama:2021xfj,Csaki:2022cyg} to derive some exact results in the limit of small $m_{3/2}$. In particular, for s-confining SQCD with $N_c \geq 3$, a perturbative analysis indicates that the vacuum exhibits $\chi$SB with the same pattern as ordinary QCD: $SU(N_f)_L \times SU(N_f)_R \rightarrow SU(N_f)_V$. This suggests an intriguing possibility that one can smoothly extrapolate from small to large $m_{3/2}$ without encountering any phase transitions and that the theory at small $m_{3/2}$ is in the same universality class as the one at large $m_{3/2}$. Unfortunately, this cannot be done in general, there are two known theories where the small and large $m_{3/2}$ theories have different vacua: SQCD with $N_c=2$ and $N_f=3$~\cite{Easa:2022chy,Csaki:2022cyg}, and a Georgi-Glashow-like $SU(5)$ theory with three generations~\cite{Bai:2021tgl}. See also~\cite{Luzio:2022ccn,Dine:2022req} for further discussion of phase transitions.

Since the original ASQCD proposal in~\cite{Murayama:2021xfj}, new results were obtained for many different kinds of gauge theories~\cite{Csaki:2021xhi,Csaki:2021aqv,Csaki:2021jax,Csaki:2021xuc,Murayama:2021rak,Kondo:2022lvu} that are consistent with smooth extrapolation from small to large $m_{3/2}$. A critical question is when such an extrapolation makes sense in general. We first make the following observation: the two theories noted above where such an extrapolation fails both have a classically conformal dynamical superpotential in the low energy theory, and they are the only s-confining theories with such a property~\cite{Csaki:1996sm,Csaki:1996zb}. Because AMSB couples to the scale anomaly, classically conformal theories only feel SUSY breaking at loop level which may be related to why those two theories are special. 

In this work, we do a more comprehensive study of s-confining ASQCD. For those theories, the dynamical superpotential can be written as follows:
\begin{equation}
W_\text{dyn} = \frac{1}{\Lambda^{N_c-2}} \det M - \widetilde{B} M B \, ,
\end{equation}
where $M$, $B$, and $\widetilde{B}$ are the low energy composites and $\Lambda$ is the confining scale described in detail in Section~\ref{sec:AMSB}. For $N_{c}=2$ the theory is classically conformal. For larger $N_c$, we have one non-renormalizable term and one classically conformal one. If we flow to low energies, the non-renormalizable operator decreases much faster, and at sufficiently low energy the theory should be described by an approximately classically conformal one in terms of only the $\widetilde{B} M B$ operator. 

Motivated by these observations, we extend the analysis of~\cite{Murayama:2021xfj,Csaki:2022cyg} for s-confining ASQCD theories. 
We expand the analysis of these models by including two and three-loop contributions to the SUSY breaking parameters to investigate the robustness of the results. The higher loop terms allow us to quantitatively analyze under what circumstances perturbation theory is a good description. We perform numerical minimization of the potential, which highlights that there are only two important vacuum configurations: s-confining with no $\chi$SB, and QCD-like $\chi$SB. 

We follow up with an analytical understanding of the QCD-like vacuum, showing that the inclusion of two and three-loops can have an impact on the phase structure. We show that the phase structure has a strong dependence on the size of the dimensionless coupling. Nevertheless, if we include the perturbativity limit, the theory is still in the QCD-like $\chi$SB as we approach the confining scale. At leading order, the phase diagram has  \textit{no} dependence on $m_{3/2}$, which could be an indication that nothing special happens when we increase $m_{3/2}$ close to the confining scale. We also highlight the special scaling of the solution with the inverse power of the non-renormalizable coupling for the  QCD-like $\chi$SB solution which is also present in the analysis done in~\cite{Murayama:2021xfj,Csaki:2022cyg}. This feature can spoil theoretical control of the solution since we need the fluctuations to be much smaller than the strong scale to justify including only the leading order Kähler potential. 

We then extend the previous analysis by including next-to-leading order Kähler corrections. As the theory approaches the confining scale, it starts to probe more Kähler terms that have information on the UV theory. Since we expect that for $N_{c}>2$ there is no phase transition as we cross the confining scale,  Kähler corrections should not spoil the $\chi$SB. However, for $N_{c}=2$, which at leading order has no $\chi$SB, the inclusion of higher order Kähler terms could give hints of the existence of $\chi$SB for large $m_{3/2}$.

 For $N_{c}=3$ we notice that the first order Kähler corrections do introduce an $m_{3/2}$ dependence. More importantly, the solutions which were runaway at small coupling become stabilized to the origin, which signals that the phase transition occurs for finite small couplings inside the regime of applicability of AMSB. This feature is different from the expectation from the initial analysis done in~\cite{Murayama:2021xfj,Csaki:2022cyg}. The expectation is still that the theory is in the $\chi$SB as we approach the confining scale.  We also apply the first-order Kähler corrections for $N_{c}=2$ which now exhibits a region of the parameter space with $\chi$SB. The phase diagram now shows hints of the dependence on $\Lambda$, as the $\chi$SB region occurs for large values of the Wilson coefficient. There is also a large portion of the $\chi$SB solution which lies outside of the EFT regime which points again to the sensitivity to what happens at $\Lambda$. To have more conclusive results it is necessary to have some theoretical control on the hierarchy of the Kähler coefficients, which is no trivial task for confining theories.

The remainder of this paper is organized as follows. In Section~\ref{sec:AMSB}, we briefly review the anomaly mediation SUSY breaking and the important results for s-confining SQCD. In Section~\ref{sec:phase} we perform the numerical and analytical analysis of s-confining ASQCD. In Section~\ref{sec:Khaller3} we analyze the first order Kähler corrections for $N_{c}=3$. In Section~\ref{sec:Khaller2}, we explore the special case of $N_{c}=2$ at leading and next-to-leading order in the Kähler potential. We conclude in Section~\ref{sec:conc}.

\section{Anomaly mediation and s-confined SQCD}\label{sec:AMSB}

The approach we use to break supersymmetry is anomaly-mediated SUSY breaking (AMSB)~\cite{Pomarol:1999ie,Randall:1998uk,Giudice:1998xp,Jung:2009dg}. The special characteristic of AMSB is that the SUSY breaking is directly connected with conformal violation of the theory. This gives us an upper hand in describing confining theories since we know how to write the conformal violation in terms of the low-energy degrees of freedom, and thus we can describe SUSY breaking directly at low energies.  This is the  UV insensitivity of AMSB, which was recently motivated to describe the dynamics of QCD-like theories~\cite{Murayama:2021xfj,Csaki:2022cyg}. 

We can introduce AMSB using the Weyl compensator field, $\Phi = 1 + \theta^{2} m_{3/2}$.  The SUSY-breaking Lagrangian can be calculated as:
\begin{align}
\mathcal{L}_{\text {\bcancel{susy}}}=\int d^4 \theta \Phi^* \Phi K+\int d^2 \theta \Phi^3 W+ \text{ h.c.} \, ,
\end{align}
where $K$ is the Kähler potential of the theory, and $W$ is the superpotential of the chiral superfield $\varphi_{i}$. If the theory is conformal, then we can re-scale the fields such that we remove $\Phi$ from the Lagrangian. We can write out the contribution from conformal breaking at the tree-level:
\begin{align}
V_{\text {\bcancel{tree}}} & =m_{3/2}\left(\partial_i W g^{i j^*} \partial_j^* K-3 W\right)+ \text{h.c.} +|m_{3/2}|^{2}\left(\partial_i K g^{i j^*} \partial_j^* K-K\right)  \, ,
\end{align}
where $g^{ij^*}$ is the inverse Kähler metric $g_{ij^*}=\partial_{i}\partial_{j^*}K$. In the limit where the SUSY breaking scale is small compared to the dynamical scale, $m_{3/2} \ll \Lambda$, we can assume that the  Kähler potential is canonical. We explore the effects of higher order Kähler corrections in Section~\ref{sec:Khaller3}. If the Kähler potential is canonical we can write the tree-level SUSY breaking effect as:
\begin{align}
V_{\text {\bcancel{tree} }}=m_{3/2}\left(\varphi_i \frac{\partial W}{\partial \varphi_i}-3 W\right) + \text{ h.c.} \, .
\end{align}
These effects vanish in classically scale invariant theories. 

There are also loop-level supersymmetry breaking effects from the superconformal anomaly.  They lead to tri-linear couplings, scalar masses, and gaugino masses\footnote{We define the anomalous dimension, its derivative, and the $\beta$ function as: $\gamma_i=\mu \frac{d}{d \mu} \ln Z_i, \dot{\gamma}=\mu \frac{d}{d \mu} \gamma_i$, and $\beta\left(g^2\right)=\mu \frac{d}{d \mu} g^2$. }
\begin{align}
A_{i j k} & =-\frac{1}{2}\left(\gamma_i+\gamma_j+\gamma_k\right) m_{3/2} \, , \\
\label{eq:softmass} m_i^2 & =-\frac{1}{4} \dot{\gamma}_i |m_{3/2}|^2 \, , \\ 
\label{eq:gauginomass} m_\lambda & =-\frac{\beta\left(g^2\right)}{2 g^2} m_{3/2} \, .
\end{align}

We are now ready to explore the ASMB version of s-confining SQCD. We are working with $SU(N_{c})$ gauge theories with $N_{f}=N_{c}+1$ vectorlike fundamental flavors. We focus first on $N_{c}>2$, while we extend the analysis of $N_{c}=2$ in Section~\ref{sec:Khaller2}.  
In the UV, the theory is described by quarks $q_{\alpha i}$ and anti-quarks $\bar{q}^{\alpha j}$, where we use Greek letters for gauge indices and Latin letters for flavor indices. In the absence of a superpotential, this theory has an $SU(N_f)_L\times SU(N_f)_R\times U(1)_B\times U(1)_R$ global symmetry. Once coupled to SUSY breaking, the $U(1)_R$ will be explicitly broken. 

This theory confines at a scale $\Lambda$, and at low energy it is described by gauge invariant composites~\cite{Seiberg:1994pq}
\begin{eqnarray} 
B^i &=& \varepsilon^{\alpha_1 ... \alpha_{N_c}} \varepsilon^{i_1 ... i_{N_c}i} \, q_{\alpha_1 i_1}\, ... \, q_{\alpha_{N_c} i_{N_c}} \nonumber\\
\widetilde{B}_i &=& \varepsilon_{\alpha_1 ... \alpha_{N_c}} \varepsilon_{i_1 ... i_{N_c}i} \, \bar{q}^{\alpha_1 i_1}\, ... \, \bar{q}^{\alpha_{N_c} i_{N_c}} \nonumber\\
M^i_j &=& q_{\alpha j} \bar{q}^{\alpha i} \, .
\end{eqnarray}
The theory is invariant under a discrete charge conjugation symmetry that exchanges $q$ and $\bar{q}^{T}$ superfields. This symmetry is preserved at the origin of the Moduli space and in the IR the theory remains invariant under $B \rightarrow \widetilde{B}^{T}$ and $M\rightarrow M^T$. At energies small compared to $\Lambda$, we can take a canonical Kähler potential for the Baryon and Meson fields ($B_{i}, \widetilde{B}_{i}, M_{ij}$). There is also a dynamically generated superpotential:
\begin{align}
W=\lambda \frac{\operatorname{det} M}{\Lambda^{N_c-2}}-\kappa \widetilde{B} M B \, .
\label{eq:superpotential}
\end{align}

We can use Naïve Dimensional Analysis (NDA)~\cite{Weinberg:1978kz,Luty:1997fk,Cohen:1997rt} to estimate that these couplings should be $\kappa \approx 4\pi$ and $ \lambda \approx (4\pi)^{N_{c}-1}$ at the strong scale. We can then employ RG running to lower energies until these couplings are small enough that we can perform perturbative calculations. Without coupling to SUSY breaking, this theory has a rich moduli space, and there is no chiral symmetry breaking at the origin of moduli space. This makes it an s-confining theory. 

We now couple the theory to AMSB. In the UV, there is no superpotential so the theory is classically conformal, and therefore the leading SUSY breaking appears at one-loop. These effects can be calculated in terms of the gauge coupling $\beta$-function and the anomalous dimensions of the quarks:
\begin{align}
\beta(g^{2}) &= - (2N_{c}-1) \frac{g^{4}}{8\pi^{2}} \, , \\
\gamma_{q}  &= \frac{N_{c}^{2}-1}{2N_{c}} \frac{g^{2}}{4\pi^{2}} \, .
\end{align}
The masses for the gaugino and squarks can then be calculated from Eqs.~\eqref{eq:softmass} and~\eqref{eq:gauginomass}:
\begin{align}
m_\lambda & =(2N_{c}-1) \frac{g^{2}}{16\pi^{2}} m_{3/2} \, , \\
\label{eq:UVmass} m_{q}^{2} &=\frac{(N_{c}^{2}-1)(2N_{c}-1)}{N_{c}} \frac{g^{4}}{64\pi^{4}} m_{3/2}^{2}\, . 
\end{align}
The masses are positive, and this conclusion remains true with higher-order corrections as long as the perturbation theory is valid. 
The fields are thus driven to scales below $\Lambda$ where the theory is better described by the composite fields.  Because of the UV insensitivity of AMSB, we do not need to follow the operators from the UV, we can simply use the AMSB results on the low energy theory of composite states.

In the IR theory, we have the superpotential from Eq.~\eqref{eq:superpotential} which has classical conformal violation for $N_{c}>2$. Once we couple the theory to AMSB this generates a tree-level interaction for the scalar components of the meson field:
\begin{align}
V_{\bcancel{\text{tree}}} = \left(N_{c}-2 \right)\frac{\lambda}{\Lambda^{N_{c}-2}}m_{3/2}\det M + \text{h.c.} \, .
\end{align}
For the loop-induced contributions, our computational approach for the low energy theory is described in Appendix~\ref{app:not}. Only the renormalizable coupling $\kappa$ contributes to the anomalous dimensions of the fields. Non-renormalizable couplings do not contribute to the renormalization of dimensionless objects like anomalous dimension functions~\cite{Weinberg:1973xwm}.
The SUSY non-renormalization theorems~\cite{Salam:1974jj,Grisaru:1979wc} imply that we can write the $\beta$-function for both couplings in dimensional reduction (DR) in terms of the anomalous dimension functions as:
\begin{align}
\beta(\kappa^{2})&= - \left(\gamma_{M} + 2 \gamma_{B}\right) \kappa^{2} \, , \\ \label{eq:betalambda}
\beta(\lambda^{2})&= - \left(N_{f}\gamma_{M}\right) \lambda^{2} \, ,
\end{align}
where we have used the discrete charge conjugation symmetry to set $\gamma_{\widetilde{B}}= \gamma_B$. We can write the anomalous dimension for the chiral fields up to three-loops as:
\begin{align}
\gamma_{M} &=\frac{1}{16\pi^2} \gamma_{M}^{(1)}+ \frac{1}{\left(16\pi^2\right)^2}\gamma_{M}^{(2)} + \frac{1}{\left(16\pi^2\right)^3} \gamma_{M}^{(3)} \, , \\
\gamma_{B} &= \frac{1}{16\pi^2}\gamma_{B}^{(1)}+ \frac{1}{\left(16\pi^2\right)^2}\gamma_{B}^{(2)} +\frac{1}{\left(16\pi^2\right)^3} \gamma_{B}^{(3)} \, ,
\end{align}
\begin{align}
\gamma_{M}^{(1)} = -2 |\kappa|^{2}\ \, , \, 
\gamma_{M}^{(2)} = 4 N_{f} |\kappa|^{4} \, , \,  
\gamma_{M}^{(3)} = - 2\left( N_{f} (4 + N_{f}) + 6 \zeta (3)\right)|\kappa|^{6}   \, ,
\end{align}
\begin{align}
\gamma_{B}^{(1)} =  - 2 N_{f}|\kappa|^{2} \, , \, 
\gamma_{B}^{(2)} = 2 N_{f} (N_{f}+1) |\kappa|^{4} \, , \,  
\gamma_{B}^{(3)} = - 2 N_{f} (-1 + N_{f} (5 + N_{f}) + 6 \zeta (3))|\kappa|^{6} \, .
\end{align}

Plugging into Eq.~\eqref{eq:softmass}, we get that the leading order soft masses are given by
\begin{eqnarray}
m_M^2 &=& \frac{2N_f+1}{128\pi^4}|\kappa|^4 |m_{3/2}|^2 \, , \\
m_B^2 = m_{\widetilde{B}}^2 &=& \frac{N_f (2N_f+1)}{256\pi^4}|\kappa|^4 |m_{3/2}|^2
\end{eqnarray}
which are of two-loop order and \textit{positive}. As long perturbation theory is reliable and the one-loop values of the anomalous dimension functions are larger than the higher order ones, then the origin of moduli space will be a stable or metastable vacuum without chiral symmetry breaking. This is not what we expect from non-SUSY QCD. There is also a one-loop $A$-term, and the origin will not be the true minimum if this trilinear interaction is large enough to generate a deeper vacuum.

We now explore the potential in detail. 
For $N_{c}>2$, we can use flavor rotations to write the distinct vacuum directions as~\cite{Csaki:2022cyg}:
\begin{align}
B=\left(\begin{array}{c} \label{eq:ansatz}
b \\
0 \\
\vdots \\
0
\end{array}\right) \, , \,  \widetilde{B}=\left(\begin{array}{c}
\bar{b} \\
0 \\
\vdots \\
0
\end{array}\right) \, , \,  M=\left(\begin{array}{llll}
x & & & \\
& v & & \\
& & \ddots & \\
& & & v
\end{array}\right) \, . 
\end{align}
Since we assume that $m_{3/2}$ is real, it is sufficient to look for minima with all fields real. The potential is minimized when $b = \bar{b}$ since the potential is symmetric in relation to them, so we can look for the minima in terms of three directions $(v, x, b)$. We can write the potential for the ASCQD with $N_{f} = N_{c} + 1$ as:
\begin{align}\label{eq:pott}
V_{\text{ASQCD}} & = V_{\text{tree} + \text{\bcancel{tree}}} + V_{\text{\bcancel{loop}}}  \, ,
\end{align}
where the slashes denote terms arising from SUSY breaking. Then we have
\begin{align}   \label{eq:potltree}
V_{\text{tree} + \text{\bcancel{tree}}} &=  \left(\lambda  \Lambda ^{2-N_{c}} v^{N_{c}}-b^2 \kappa \right)^2 + 2 b^2 \kappa ^2 x^2 + 2 \lambda  m_{3/2} (N_{c}-2) x \Lambda ^{2-N_{c}} v^{N_{c}}+\lambda ^2 N_{c} x^2 \Lambda ^{4-2 N_{c}} v^{2 (N_{c}-1)} \, , \\ \label{eq:potloop}
V_{\text{\bcancel{loop}}} &= -\frac{1}{2} b^2  \dot{\gamma}_{B} m_{3/2}^{2}-\frac{1}{2} b^2 x \kappa m_{3/2}(-\gamma_{M}-2\gamma_{B}) - \frac{1}{4} \dot{\gamma}_{M}m_{3/2}^{2} \left( N_{c} v^{2}+x^{2}\right) \, .
\end{align}
The analysis of this potential done in~\cite{Murayama:2021xfj,Csaki:2022cyg} performed the minimization with only $ V_{\text{tree} + \text{\bcancel{tree}}}$ and then checked the stability to corrections from $V_{\text{\bcancel{loop}}}$. Using this approximation, they identified two possible non-trivial vacuum configurations: The baryon conserving and the baryon breaking minimum. We show that ignoring $V_{\text{\bcancel{loop}}} $ is not a valid approximation even in the small  $m_{3/2}$ limit for the baryon conserving direction. The quadratic term plays an important role in the existence of this minimum. The identification of the minima stays correct for $\kappa \rightarrow 0$ ($ \dot{\gamma}_{M}=0$) which reproduces the approximations in~\cite{Murayama:2021xfj,Csaki:2022cyg}.

\section{Analysis of the phase diagram up to three-loops} \label{sec:phase}

Before delving into the analysis of this theory, it is crucial to address the parameterization of its phase diagram using $\kappa$ and $\lambda$. This parameterization arises due to a lack of information regarding the matching with the UV theory. Essentially, this theory is described by a single parameter, denoted as $m_{3/2}/\Lambda$. For a given energy and a specific value of $m_{3/2}$, only one point in the ($\kappa, \lambda$) plane is realized. However, since we are unaware of which specific point that is, we explore the entire ($\kappa, \lambda$) plane for different values of $m_{3/2}$.

Even though we do not know the value of the couplings at the strong scale, we know how they will run down to the IR, at least within the perturbativity regime. Since we have dimensionful parameters, it is better to define the running considering the total dimension of the couplings. In this case, we remove every dimension from the dimensionful coupling and substitute with powers of the RG running parameter $\mu$. The $\beta$-function for these couplings will have then the loop suppressed term from Eq.~\eqref{eq:betalambda}, but also will have the leading term from its dimension, which in the case of $\lambda$ is $(N_{c}-2)$.

Another important thing to notice is that we have a natural scale to probe this theory. Since the SUSY breaking term is massive, we naturally probe the theory choosing $\mu = m_{3/2}$. So, in this case, we run the theory from $\Lambda$ to $m_{3/2}$. The running near the scale $\Lambda$ is outside of the perturbativity regime. However, we can look from the other direction and start flowing from $\mu=m_{3/2}$ to $\Lambda$ and ask what values at $\mu = m_{3/2}$ would generate the NDA expectation at $\mu =\Lambda$. 

We now turn to minimizing the potential as a function of free parameters $(\kappa, \lambda)$. It is not possible to obtain a simple analytical solution for the equations of motion coming from Eq.~\eqref{eq:pott} when $b\neq 0$. We, therefore, resort to numerical minimization of the potential for the most general case. As noted above, we keep terms proportional to $m_{3/2}^{2}$. This section is then organized as follows: we first perform numerical scans of the full potential. Then, using the additional information from the numerical scan, we follow up with an analytical understanding of the phase diagram. After the leading behavior is understood, we explore in the next sections the stability of these solutions once we include higher order Kähler terms for the $N_{c}=2$ and $N_{c}=3$ cases.

\subsection{Numerical scan of the phase diagram}

To investigate the vacuum structure, we performed numerical scans for local minima across different values of $N_{c}$ (ranging from 3 to 8), loops (one, two, and three), and $m_{3/2}$ ($0.001\Lambda$, $0.01\Lambda$ and $0.1\Lambda$). For each minimization, we determined the global minimum and identified the second local minimum, if it existed. For each solution, we also checked how far away from the origin it is. As our effective field theory (EFT) only works at energies below the confining scale $\Lambda$, we consider configurations with any of $(x,v,b) > \Lambda$ to be outside of the validity of the EFT. 

 The first important result that we obtained from the scans is that there are only two possible global minima: 
 \begin{itemize}
 \item The s-confining vacuum at the origin: $b=v=x=0$.
 \item The QCD-like $\chi$SB vacuum with $v= \pm x \neq 0$\footnote{Since we are fixing $\lambda$ and $\kappa$ to be positive, in the scans we are allowing the vevs to be negative to explore the full range of the parameter space. In our convention $v= - x \neq 0$ is the QCD-like $\chi$SB minimum. } and $b=0$.
 \end{itemize}
 \begin{figure*}[t!]
   \resizebox{0.35\linewidth}{!}{\includegraphics{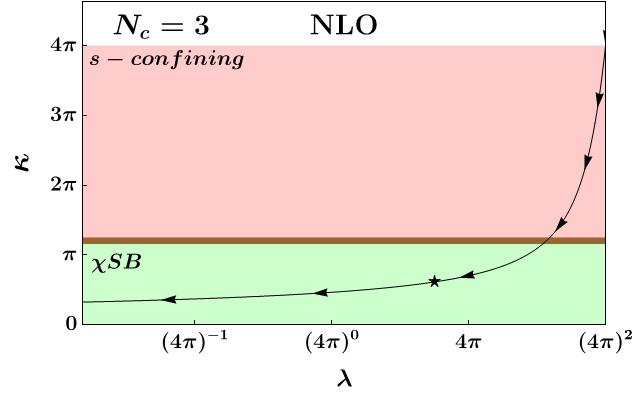}}
   \resizebox{0.35\linewidth}{!}{\includegraphics{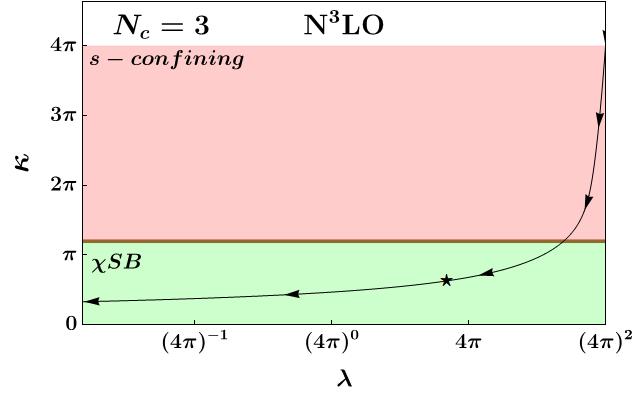}}
   \resizebox{0.35\linewidth}{!}{\includegraphics{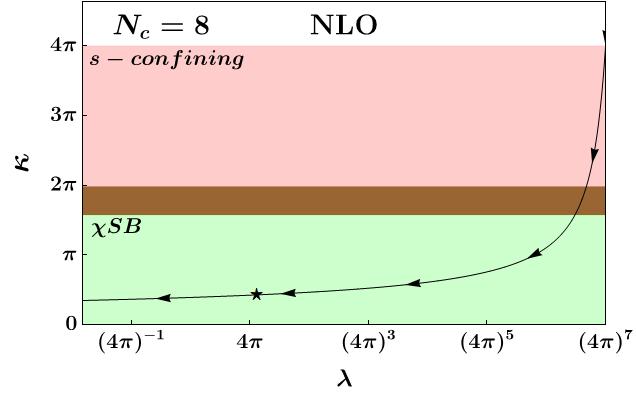}}
   \resizebox{0.35\linewidth}{!}{\includegraphics{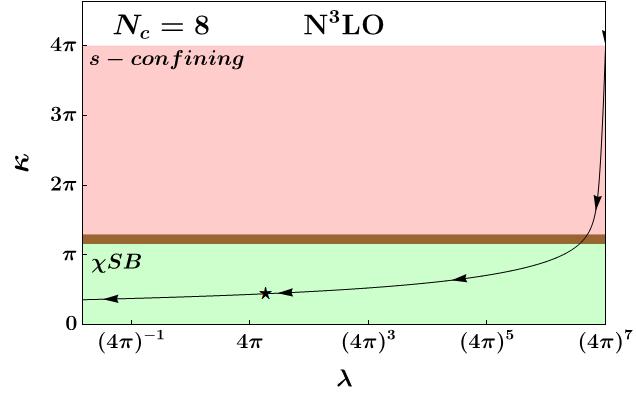}}
     \caption{ \label{fig:NUMPHASE} Phase diagram of $\kappa$ vs $\lambda$ for $N_{c} = 3$ (top) and $N_{c} = 8$ (bottom) at one and three-loops (left to right). The two-loop result has $\chi$SB for all values of $\kappa$ as explained in the text. These results are the boundary obtained from the numerical minimizations and they do not depend on $m_{3/2}$. In red, at the top, we have the region with unbroken symmetry. The brown middle region also has its minimum at the origin, but it contains a local minimum with QCD-like $\chi$SB. In the green region, the QCD-like $\chi$SB vacuum is the global minimum. The line indicates the RG flow from the NDA value at $\Lambda$. The star indicates the value of the couplings which starts at NDA and ends at $m_{3/2} = 0.1 \Lambda$.}
\end{figure*}

The phase diagram is shown in Figure~\ref{fig:NUMPHASE}, for $N_{c}=3$ and $N_{c}=8$.   
We see that the phase diagram is independent of $\lambda$, the coefficient of the non-renormalizable operator in the superpotential (c.f.~Eq.~\eqref{eq:superpotential}). We also see that at small values of $\kappa$, we get a $\chi$SB vacuum which is analogous to what happens in non-SUSY QCD and consistent with the results of~\cite{Murayama:2021xfj,Csaki:2022cyg}. For larger values of $\kappa$, the origin is the minimum. We highlight that at the boundary of the two vacua, there is a small region of parameter space where the QCD-like $\chi$SB vacuum exists, but it has positive vacuum energy and the origin is the global minimum. This region is shown in brown in Figure~\ref{fig:NUMPHASE}. The two-loop result is substantially different from the one/three-loop because of the sign change in the anomalous dimension. At the two-loop level, it is possible to change the sign of $\dot{\gamma}_{M}$, and this occurs before the potential location of the brown boundary, which in turn means that we never change the phase to s-confining. The change in sign of $\dot{\gamma}_{M}$ is a clear indication that we are moving outside the regime of validity of perturbation theory. 

In Figure~\ref{fig:NUMPHASE} we applied the NDA value as the upper bound on $\kappa$, but since we have the higher-loop information we can do better. As we include both two and three-loop contributions to the anomalous dimensions, we can then compute a perturbativity bound by restricting that a given contribution is always larger than the next loop order.
 This pertubativity bound depends on which observable we are interested in. Since the QCD-like $\chi$SB vacuum is mostly sensitive to $\dot{\gamma}_{M}$, we use this as our important observable. We then compare different orders of the perturbation theory and whenever equality occurs in any of these comparisons, we have a guarantee that perturbation theory ceases to be a good description.  The strongest bound we obtain is from comparing the one-loop and two-loop coefficients and gives us the following perturbativity upper bound on $\kappa$:
\begin{align}
\kappa < 2\pi  \sqrt{\frac{4 N_{c}+6}{5 N_{c}^2+14 N_{c}+9}} \, .
\end{align}

\begin{figure*}[t!]
   \resizebox{0.45\linewidth}{!}{\includegraphics{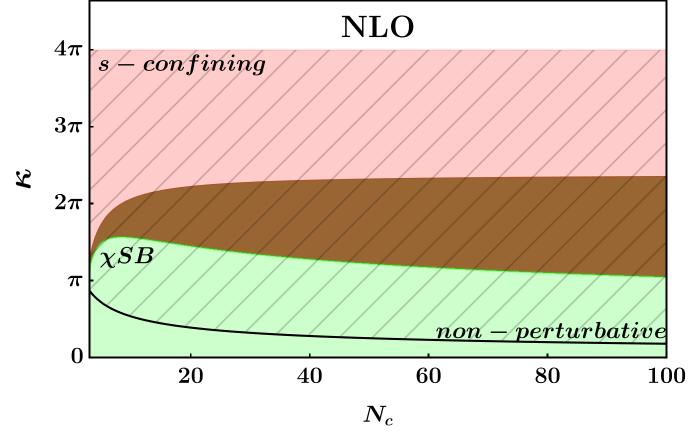}}
   \resizebox{0.45\linewidth}{!}{\includegraphics{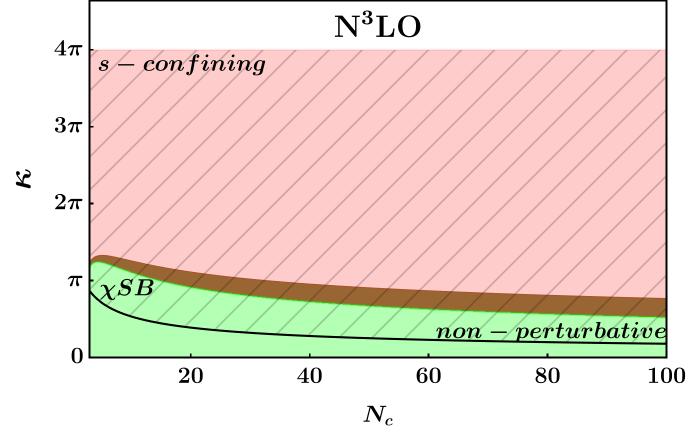}}
     \caption{ \label{fig:kvsNC} Phase diagram of $\kappa$ vs $N_{c}$, for one and three-loops respectively. The phase diagram does not depend on $m_{3/2}$. We use the same colors as in Figure~\ref{fig:NUMPHASE}.  The hashed region is outside the regime of perturbativity which we set by looking at the size of the coefficients of $\dot{\gamma}_{M}$. 
     }
\end{figure*}

We can visualize the perturbativity bound and the phase diagram as a function of $N_{c}$ in Figure~\ref{fig:kvsNC}.  The phase diagram for arbitrary $N_{c}$ in Figure~\ref{fig:kvsNC} is obtained from the analytic understanding of the QCD-like vacuum in the next subsection. The phase transition from the QCD-like $\chi$SB to the s-confining vacuum occurs outside the regime of validity of perturbation theory. This means that the s-confining vacuum may not be realized in the full theory. 

Furthermore, the possibility that the QCD-like $\chi$SB vacuum in the SUSY and non-SUSY theories are simply connected in theory space as $m_{3/2}$ is varied remains quite likely. Additionally, the pertubativity boundary for the two-loop result is exactly when $\dot{\gamma}_{M}$ changes sign, which is the reason why there is no s-confining regime at the two-loop level. This boundary is recovered when we include the three-loop result and there is no more sign change even when we cross this perturbativity regime. 

We can also explore how far we need to flow from $\Lambda$ starting at the NDA values of the couplings to reach the perturbative boundary. This result is an extrapolation of the regime of applicability of these results, as is the NDA in nature, but can be useful to have an expected value to consider $m_{3/2}$ small. At one-loop, we can exactly solve the RG flow of the couplings and we reach the perturbative boundary from NDA for:
\begin{align}
\log \frac{m_{3/2}}{\Lambda} \approx -\frac{15+26N_{c}+10N_{c}^{2}}{2\left( 3 + 2N_{c} \right)^{2}} \, .
\end{align}
For $N_{c}=3$ we have $\frac{m_{3/2}}{\Lambda}  \approx 0.32$ and the value for the ratio slowly gets smaller as we increase $N_{c}$ until the saturation point at $N_{c} \rightarrow \infty$ where $\frac{m_{3/2}}{\Lambda}  \approx 0.28$. We can see that flowing from NDA we would be inside the perturbativity region for the values of $m_{3/2}$ that we explore. In Figure~\ref{fig:NUMPHASE} the star indicates the value of the couplings assuming NDA values and flowing down up to $m_{3/2}=0.1\Lambda$.

While not shown in Figures~\ref{fig:NUMPHASE} and~\ref{fig:kvsNC}, our numerical analysis showed that for small enough $\lambda$, the $\chi$SB solution becomes a runaway outside of the validity of the EFT. This signals that the size of the vev scales with a negative power of $\lambda$. This is indeed what we find analytically, and this result still holds in the $\kappa \rightarrow 0 $ limit. In the next subsection, we analytically explore the QCD-like $\chi$SB vacuum to derive an expression for both boundaries. The phase diagram is also independent of $m_{3/2}$, but thus far we have ignored higher order Kähler corrections whose effects are proportional to $m_{3/2}/\Lambda$. Those corrections will be explored in Section~\ref{sec:Khaller3}.

We find there is no possibility of having a global vacuum that breaks baryon number. This is also what happens in the limit of large SUSY breaking due to the Vafa-Witten theorem~\cite{Vafa:1983tf}.  There is, however, a region of parameter space that has a local minimum that breaks baryon number and has negative vacuum energy, but that minimum is always higher than the baryon preserving minimum. We discuss this further in Appendix~\ref{app:secondMIN}.

\subsection{Analytical results for the baryon conserving chiral breaking vacuum}

Motivated by our numerical analysis, we can analytically explore the QCD-like $\chi$SB vacuum ($b=0$, $x=-v$) in more detail. In that case, the potential simplifies to:
\begin{align}
V = -\frac{1}{4} \dot{\gamma}_{M} m_{3/2}^{2} (N_{c}+1) v^2 - 2 \frac{\lambda}{\Lambda^{N_{c}-2}} m_{3/2} (N_{c}-2) v^{N_{c}+1} + \left(\frac{\lambda}{\Lambda^{N_{c}-2}}\right) ^2 (N_{c}+1) v^{2 N_{c}} \, . 
\end{align}
Notice that naively we would drop the $m_{3/2}^{2}$ term since we are interested in the region where $m_{3/2}$ is small. However, we can perform the following rescaling $v \rightarrow v \left( m_{3/2} \right)^{1/(N_{c}-1)}$ such that every term has the same contribution from $m_{3/2}$ and thus the loop term is as important as the others. Keeping all the terms we can still find a solution for the minimization:
\begin{align} \label{eq:MIN} \nonumber
v_{\text{MIN}} &= \xi^{ 1/(N_{c}-1)} \, \\
\xi &= \frac{m_{3/2}}{2 \Lambda^{N_{c}-2} N_{c} \lambda }\left(N_{c}-2 + \sqrt{4 + N_{c}\left(N_{c}-4+ \dot{\gamma}_{M} \right) } \right) \, .
\end{align}
 This solution reduces to the solution described in~\cite{Murayama:2021xfj} once we set $\dot{\gamma}_{M} = 0 $ or equivalently $\kappa = 0$. The phase structure can then be potentially more complex than what was described in~\cite{Murayama:2021xfj}. The existence of this solution now depends on $\dot{\gamma_{M}}$. As we saw in Figure~\ref{fig:kvsNC}, the potential phase transition to the s-confining phase is well above the perturbativity regime and thus we expect the theory to be in the symmetry breaking phase as we approach the strong scale.

 Another important feature to point out that was previously overlooked is the dangerous $1/\lambda$ scaling of this solution. The theoretical control of the calculation relies on the fluctuations being smaller than the strong scale. This means that at small enough $\lambda$, the size of the solution will be large enough that we cannot trust the calculations. If the vevs goes beyond $\Lambda$, it becomes necessary to reformulate the theory using UV quark and gaugino superfields. However, coupling small AMSB to the UV theory results in the stabilization of squarks at the origin, characterized by positive masses (c.f.~Eq.~\eqref{eq:UVmass}). This scenario corresponds to a runaway behavior within the UV theory, which remains valid only at high vevs. Since both theories converge towards $\Lambda$, this indicates that the theory stabilizes in the vicinity of $\Lambda$. We can explore what values of $\lambda$ generate these solutions by exploring when $v_{\text{MIN}}=\Lambda$. 
 
 Using only the tree-level terms in the solutions of Eq.~\eqref{eq:MIN} we can estimate a lower bound on $\lambda$:
 \begin{align}
 \lambda_{\text{runaway}} \approx \frac{m_{3/2}}{\Lambda} \left(\frac{N_{c}-2}{N_{c}} \right) \, .
 \end{align}
 The contribution from $\dot{\gamma}_{M}$ generates a small dependence on $\kappa$. The parameter space that this occurs can be, for example, higher than $\lambda = 0.01$ for values of $m_{3/2}=0.1\Lambda$. We can explore how far away from $\Lambda$ we need to flow to reach the runaway solution assuming the initial NDA value. This will give us an estimate on the range of $m_{3/2}$ which we trust the calculation. The leading contribution for the running of $\lambda$ comes from its dimension:
 \begin{align}
 \lambda(m_{3/2}) \approx \left(4\pi \right)^{N_{c}-1} \left(\frac{m_{3/2}}{\Lambda} \right)^{N_{c}-2} \, .
 \end{align}
 We can see that for $N_{c}=3$ we never reach the runaway solution, while for $N_{c}=4$ we reach the runaway for $m_{3/2}/\Lambda \approx  2 \times 10^{-4}$. As we increase $N_{c}$ the values for $m_{3/2}$ increases, reaching  $m_{3/2}/\Lambda \approx  1/(4\pi)$ for $N_{c} \rightarrow \infty$.
 
Now, let us understand boundaries in Figure~\ref{fig:NUMPHASE} and~\ref{fig:kvsNC}. From the solution in Eq.~\eqref{eq:MIN}, we can understand the first boundary between the s-confining and the QCD-like $\chi$SB vacuum. If the square root in Eq.~\eqref{eq:MIN} becomes negative this solution becomes imaginary and ceases to exist, since we are assuming everything real. This gives a condition for $\dot{\gamma}_{M}$:
\begin{align}
\dot{\gamma}_{M} \geq - \frac{(N_{c}-2)^{2}}{N_{c}}\, .
\end{align}
Given that $\dot{\gamma}_{M}$ is negative this gives a non-trivial condition for $\kappa$. For example, at one-loop, we have the $\chi$SB solution only when:
\begin{align}
0 < \kappa < 2 \sqrt{2} \pi  \left( \frac{(N_{c}-2)^2}{N_{c} (2 N_{c}+3)}\right)^{1/4} \, .  
\end{align}
This gives the boundary between the s-confining and the QCD-like $\chi$SB solution. The behavior of the one-loop solution is significantly different from the two/three-loop. The boundary for one-loop increases and saturates for large $N_{c}$ at $\kappa = 2^{5/4}\pi$. At two-loop, $\dot{\gamma}_{M}$ changes sign and we never have a transition, while at three-loops we recover this boundary, but now the boundary goes to zero as we increase $N_{c}$.

We can also describe the lower boundary where the QCD-like $\chi$SB solution becomes the global minimum by finding a solution that has zero vacuum energy:
\begin{align}
v_{0} = \left(2\lambda \Lambda^{2-N_{c}}\frac{4-2N_{c} + \sqrt{4(N_{c}-2)^{2}+(1+N_{c})^{2}\dot{\gamma}_{M}}}{m_{3/2}(1+N_{c})\dot{\gamma}_{M}} \right)^{1/(1-N_{c})} \, .
\end{align}
Then, we can check when this zero energy solution is equal to  Eq.~\eqref{eq:MIN} and write this condition in terms of $\dot{\gamma}_{M}$ which gives:
\begin{align}
\dot{\gamma}_{M} < -\frac{4 (N_{c}-2)^2}{(N_{c}+1)^2} \, .
\end{align}
We can solve for $\kappa$ to obtain the lower boundary of the brown region in Figures~\ref{fig:NUMPHASE} and~\ref{fig:kvsNC}. For example, at one-loop, we have the global QCD-like $\chi$SB vacuum at:
\begin{align}
0\leq \kappa <\frac{4 \pi  \sqrt{N_{c}-2}}{\left((N_{c}+1)^2 (2 N_{c}+3)\right)^{1/4}} \, .
\end{align}
The one-loop boundary goes to zero at $N_{c}\rightarrow \infty$ and the behavior is similar at three-loops. More importantly, the perturbativity boundary is \textit{always} lower than the s-confining region at this loop order.

From this analysis, we can see that without the perturbativity bound, naively one would expect a phase transition from the QCD-like $\chi$SB vacuum to the s-confining one to occur as we increase $m_{3/2}$. This ends up not happening because we reach the perturbativity bound earlier. It is also important to reinforce that the existence of the QCD-like $\chi$SB vacuum is independent of $m_{3/2}$ at with a leading order Kähler potential.

\section{Kähler corrections for $N_{c}=3$ with $N_{f}=4$} \label{sec:Khaller3}

Now, let us explore the next order Kähler correction for $N_{c}=3$. This case was the only one where the perturbativity limit was close to the $\chi$SB vacuum, and one could wonder if perturbations could change this picture. The form of the first correction of the Kähler potential is independent of $N_{c}$ and respects the discrete charge conjugation symmetry~\cite{Cho:1998vc}. Here we are only interested in the contributions to the Kähler potential that contribute to the scalar potential. Therefore we are left with:
\begin{align} \nonumber
\Lambda^{2}K_{6} &= \frac{c_{M_{1}}}{N_{f}^{2}}\Tr(M^{\dagger}M)^{2}+\frac{c_{M_{2}}}{N_{f}}\Tr(M^{\dagger}M M^{\dagger}M) + \frac{c_{B}}{N_{f}}\left((B^{\dagger}B)^{2}+(\widetilde{B}^{\dagger}\widetilde{B})^{2} \right)  \, \\ \nonumber
&+ \frac{c_{B\widetilde{B}}}{N_{f}}(B^{\dagger}B)(\widetilde{B}^{\dagger}\widetilde{B})+ \frac{c_{MB}}{N_{f}^{2}}\Tr(M^{\dagger}M)\left(B^{\dagger}B + \widetilde{B}^{\dagger}\widetilde{B} \right) \, \\
&+ \frac{c_{BMMB}}{N_{f}} \left( B M  M^{\dagger}B^{\dagger} + \widetilde{B} M  M^{\dagger}\widetilde{B}^{\dagger} \right)  ,
\end{align}
where we normalized the Wilson coefficients such that they are finite in the large $N_{f}$ limit. Using NDA we can expect that these coefficients are of order $ c_{O} \approx (4\pi)^{2}$ at the strong scale. Since these are higher dimensional operators they will run faster to zero than the leading parameters. Inside the perturbativity region, we should expect them to be small.

One thing to notice is that once we write the scalar potential from $K_{6}$, we get terms that scale as $c_{O}/\Lambda^{2}$, $c_{O}^{2}/\Lambda^{4}$ and $c_{O}^{3}/\Lambda^{6}$. One must then be careful in enforcing the correct truncation of the EFT since there are also non-renormalizable terms in the superpotential. From power counting alone it seems that the inclusion of these Kähler terms is required for the consistency of the series for higher values of $N_{c}$. In this work, we expand the potential in the Wilson coefficients and keep only the linear contribution. 

We again explore this potential using numerical analysis, similar to what was done in Section~\ref{sec:phase}. For simplicity we assume that all the Wilson coefficients are positive to enforce boundness from below.\footnote{We could have $c_{MB}$ and $c_{M\widetilde{B}}$ being small and negative numbers, but we do not consider this region. It is also important to note that boundness from below is not required in an EFT, and there could be an unstable local minimum in the theory at this truncation level. } We also consider that the theory does not break the charge conjugation symmetry between $B$ and $\widetilde{B}$ which means that we can still consider the minimization in the same directions as before. This simplifies the analysis since we keep only three non-trivial directions, $(b, v, x)$, defined in Eq.~\eqref{eq:ansatz}.

We conduct numerical investigations on the impact of $K_6$ for $N_c=3$ at one-loop. Since $K_6$ introduces six extra parameters, the resulting phase diagram in the $\kappa$ vs.~$\lambda$ plane becomes just a slice of the full phase diagram. To mitigate this problem, we avoid showing different phases in the same box so we can still distinguish between the various regions. We investigated the parameter space with this in mind, exploring different hierarchies by using different samples. 

\begin{figure*}[t!]
    \resizebox{0.32\linewidth}{!}{\includegraphics{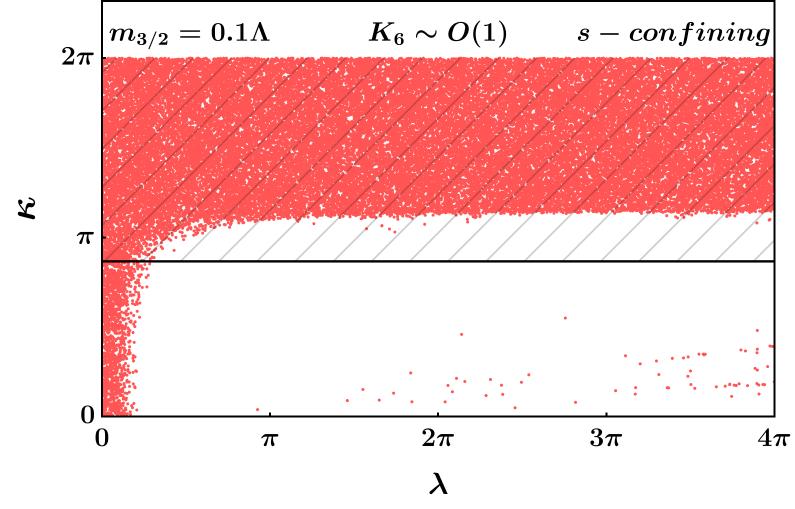}}
    \resizebox{0.32\linewidth}{!}{\includegraphics{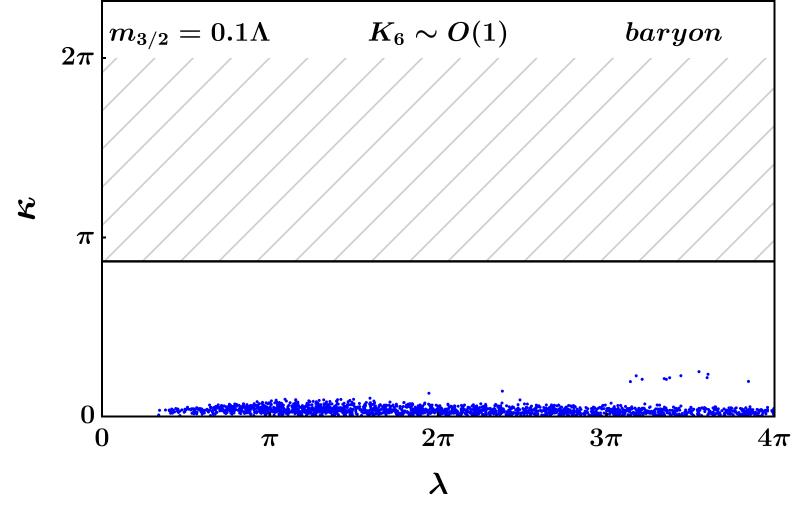}}
    \resizebox{0.32\linewidth}{!}{\includegraphics{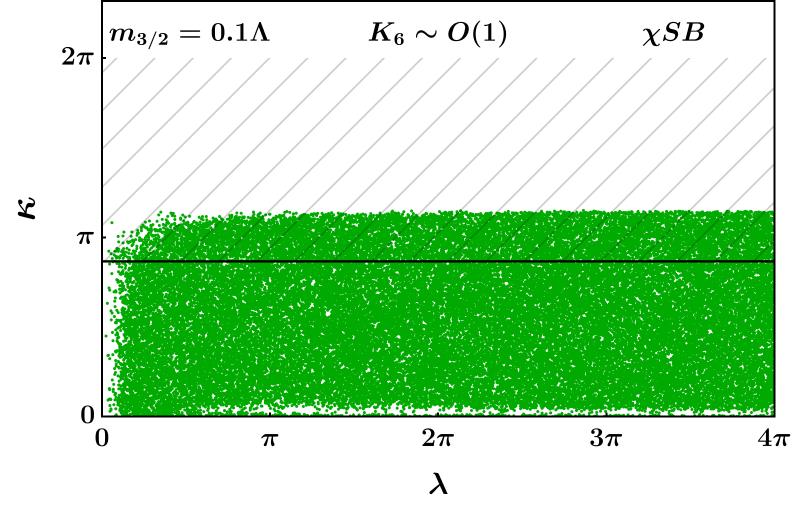}}
     \caption{ \label{fig:KHASCAN}Phase diagram for the global minimum in the $\kappa$ vs $\lambda$ plane for a random scan of the coefficients of $K_{6}$ for fixed $m_{3/2}=0.1\Lambda$. The different phases are separated into different panels for better visualization, we have the s-confining vacuum on the left, the broken baryon vacuum in the middle, and the QCD-like $\chi$SB vacuum on the right. The hashed region is the perturbativity bound on $\kappa$.}
\end{figure*}

In Figure~\ref{fig:KHASCAN} we can see the numerical scan considering a flat prior in all the parameters for $m_{3/2}=0.1\Lambda$ while restricting the Wilson coefficients to be smaller than one. We also explored the same diagram for $m_{3/2}=0.001\Lambda$ which we do not show here since it is similar to Figure~\ref{fig:KHASCAN}, but with a closer appearance to Figure~\ref{fig:NUMPHASE}. This indicates that as we dial down $m_{3/2}$ we are returning to the leading result from Figure~\ref{fig:NUMPHASE}.

One interesting result of the scan is the appearance of a region where baryon number is spontaneously broken. This does not mean that the theory has a broken baryon phase since we do not know which point in the $\kappa$ vs.~$\lambda$ plane is realized. These solutions occur for small $\kappa$ and are not deformation from the second global minimum explored in Appendix~\ref{app:secondMIN}.

 One important feature that becomes difficult to visualize in these figures is the modification of the runaway solutions which were present for small $\lambda$. With the inclusion of the leading order Kähler correction, all these solutions now become s-confining, and we can see that in the small $\lambda$ region we start to populate the s-confining phase for some portion of the parameter space. This can be seen by the simultaneous increase of the s-confining solution close to the origin and the absence of $\chi$SB solutions in the same region. This is explored further in Appendix~\ref{app:secondMIN}.

We can say that from the numerical analysis, the points which live closer to the perturbative boundary were in the QCD-like $\chi$SB vacuum. We expect from NDA that the theory is in this phase, meaning that the NLO corrections did not significantly alter the theory. In general, it is difficult to draw any significant conclusion without knowing at least the expected hierarchy of the couplings. If it were possible to calculate some of the Wilson coefficients at  $\Lambda$, then the RG running could tell us how important these Kähler corrections are, and we could have a clearer picture of which vacuum the theory lives in for $m_{3/2} \lesssim \Lambda$.  A more detailed analysis with further analytical understanding would be necessary to draw any additional conclusions.

\section{s-confining SU(2) ASQCD } \label{sec:Khaller2}

\subsection{Leading Kähler results  } \label{subsec:SU2}

The SU(2) SQCD with $N_{f}=3$ is a special theory. It is one of only two theories that is s-confining and classically conformal at low energy~\cite{Csaki:1996sm,Csaki:1996zb}. This behavior is not what we expect from the non-SUSY counterpart~\cite{KOGUT1988465,Kogut:2000ek,Huang:2014xwa,PhysRevD.85.014504,Arthur:2016dir,BMW:2013fzj} which indicates that  SUSY breaking needs to have a substantial effect on the theory. Another unique feature of this theory is the enhancement of the flavor symmetry to $SU(6)$. Both the quarks and anti-quarks can be described as fundamentals, and in the low energy, the gauge invariant bound state is $V_{ij} = Q_{i}Q_{j}$, which is an anti-symmetric tensor of $SU(6)$. There is no notion of baryon or meson direction, and the most general vacuum can have the form:
\begin{align}\label{eq:su2matrix}
V = \begin{pmatrix}
0 & \xi_{1} & 0 & 0 & 0 & 0\\
-\xi_{1} & 0 & 0 & 0 & 0 & 0\\
0 & 0 & 0 & \xi_{2} & 0 & 0\\
0 & 0 & -\xi_{2} & 0 & 0 & 0\\
0 & 0 & 0 & 0 & 0 & \xi_{3}\\
0 & 0 & 0 & 0 & -\xi_{3} & 0\\
\end{pmatrix} \, .
\end{align}  
This theory is classically conformal and has the following superpotential~\cite{Seiberg:1994bz,Seiberg:1994pq}:
\begin{align}
W_{2} = \kappa \text{Pf} V \, .
\end{align}
We can write the potential including anomaly mediation in terms of $\gamma_{V}$ and $\dot{\gamma}_{V}$:
\begin{align}
V =\kappa ^2  \left( \xi_{1}^{2}\xi_{2}^{2} + \xi_{1}^{2}\xi_{3}^{2}+ \xi_{2}^{2}\xi_{3}^{2} \right) -\frac{1}{8} \dot{\gamma}_{V} m_{3/2}^2 \left(\xi_{1}^{2} +\xi_{2}^{2} +\xi_{3}^{2} \right) - \frac{3}{2}\gamma_{V} \kappa m_{3/2} \xi_{1}\xi_{2}\xi_{3} \, .
\end{align}

A sufficient condition for the origin to be the global minimum of the potential can be derived by writing the potential as the sum of squares:
\begin{align} \nonumber
V &= \left( \kappa \xi_{1}\xi_{2} - \frac{1}{4}\gamma_{V}m_{3/2} \xi_{3} \right)^{2}+ \left( \kappa \xi_{1}\xi_{3} - \frac{1}{4}\gamma_{V}m_{3/2} \xi_{2} \right)^{2}+\left( \kappa \xi_{2}\xi_{3} - \frac{1}{4}\gamma_{V}m_{3/2} \xi_{1} \right)^{2} \,   \\
   & - \frac{1}{16}m_{3/2}^{2} \left( 2\dot{\gamma}_{V} + \gamma_{V}^{2} \right) \left( \xi_{1}^{2} +\xi_{2}^{2}+\xi_{3}^{2}  \right) \, .
\end{align}
In this form, the first three terms are always positive, and the mass term is positive when:
\begin{align}\label{eq:conda}
-2\dot{\gamma}_{V} - \gamma_{V}^{2}> 0 \, .
\end{align}

The anomalous dimension and its derivative up to three-loops are given by:
\begin{align}
\gamma_{V} =  -\frac{3 \kappa ^2}{8 \pi ^2} + \frac{9 \kappa ^4}{64 \pi ^4}  - \frac{27\kappa^{6}\left( 5-2\zeta(3)\right)}{4096\pi^{6}} \, ,
\end{align}
\begin{align}
\dot{\gamma}_{V} =-\frac{27 \kappa ^4}{64 \pi ^4}+  \frac{243 \kappa ^6}{512 \pi ^6} -\frac{243 \kappa ^8 ( 9 - 2 \zeta(3))}{8192 \pi ^8} \, .
\end{align}
We can then plug these results back into Eq.~\eqref{eq:conda} to find that, at one-loop, we never have symmetry breaking. This condition gets weakened as we go to three-loops, where the no symmetry breaking region is characterized by  $\kappa \lesssim 20.7$. This value is way above the perturbativity bound, which we set by using $\dot{\gamma_{V}}$, and using the strongest bound when the one and two-loop coefficients are equal giving $\kappa \leq \frac{2\sqrt{2}}{3}\pi\approx 3$.

We can then see that loop effects cannot change the vacuum of this theory. It looks like at small $m_{3/2}$ we have no $\chi$SB, meaning that we expect the transition to happen at large $m_{3/2}$. We cannot access the information of large $m_{3/2}$ directly since we lose theoretical control of the calculation. However, we can expect that some of this information is encoded in the Kähler coefficients. One approach which we explore in this work is to probe these higher order Kähler terms. 

\subsection{Higher order Kähler corrections for SU(2) ASQCD} 

The leading correction for the Kähler potential for $N_{c}=2$ can be parametrized using the same convention as before:
\begin{align}
\Lambda^{2}K_{6}^{N_{c}=2}= \frac{c_{1}}{9}\Tr(V^{\dagger}V)^{2} + \frac{c_{2}}{3}\Tr(V^{\dagger}VV^{\dagger}V) \, .
\end{align}
To avoid problems with unbounded potential, we require the coefficients to be positive. Our numerical analysis follows a similar recipe as our calculation for $N_{c}=3$. We first do a general minimization analysis and find that, at least at this order, the only possible breaking direction is the diagonal one: $\xi_1=\xi_2=\xi_3 = \phi$ in the parameterization of Eq.~\eqref{eq:su2matrix}. This in turn simplifies the analysis as we can focus only on this direction. The scans were performed for fixed values of $m_{3/2}=(0.001\Lambda,0.01\Lambda, 0.1\Lambda)$. We perform the analysis of the scalar potential linear in the Wilson coefficients, which in this case has a direct relation with the EFT truncation up to order $1/\Lambda^{2}$. In the diagonal direction, the potential is:
\begin{align}
V(\phi) &= -\frac{3}{8}m_{3/2}^{2}\dot{\gamma}_{V} \phi^{2} - \frac{3}{2}\gamma_{V} \kappa m_{3/2}\phi^{3} +3 \kappa^{2}\phi^{4}  \\ \nonumber
&+ 14 m_{3/2}^{2}\frac{c_{12}}{\Lambda^{2}} \phi^{4} +24\frac{c_{12}}{\Lambda^{2}} \kappa m_{3/2} \phi^{5}  + 8 \frac{c_{12}}{\Lambda^{2}} \kappa^{2}\phi^{6}  \, ,
\end{align}
where we define the coupling $c_{12}= 2c_{1}+c_{2}$. We can first analyze the effects of the Kähler term in the supersymmetric theory. In this limit, we have:
\begin{align}
V_{m_{3/2} = 0 } = 3 \kappa^{2} \phi^{4}  + 8 \frac{c_{12}}{\Lambda^{2}} \kappa^{2}\phi^{6} \, .
\end{align}
This potential has no non-trivial minima for positive $c_{12}$. For negative $c_{12}$ the potential becomes unstable and the minimum will be outside the regime of validity of the EFT.

\begin{figure*}[t!]
        \resizebox{0.65\linewidth}{!}{\includegraphics{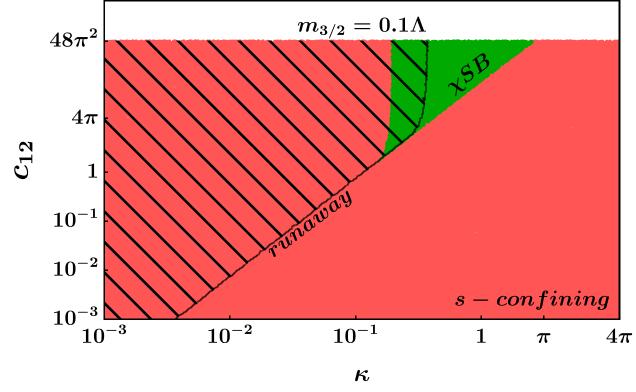}}
     \caption{ \label{fig:SU2}Phase diagram at one-loop including the first order Kähler correction for $N_{c}=2$ $N_{f}=3$ for $m_{3/2}=0.1\Lambda$. The large red region is s-confining, the green triangle on the upper right is diagonal breaking: $\xi_{1}=\xi_{2}=\xi_{3}=\phi\neq 0$ and $|\phi| < 1$, and the hashed black region are symmetry breaking points outside the EFT (We here consider a more conservative condition of $|\phi|>0.5\Lambda$). The symmetry breaking solution has a feature that as we change $m_{3/2}$, we preserve the shape of this curve by rescaling $\kappa \rightarrow \kappa \, m_{3/2}/(0.1 \Lambda)$, obtaining then the phase diagram for any $m_{3/2}$. }
\end{figure*}

Now, we turn on $m_{3/2}$, and we can see the consequence from the inclusion of the next-to-leading order Kähler correction for s-confining $SU(2)$ in Figure~\ref{fig:SU2}. In green we have symmetry breaking, and red points are s-confining. The hashed black region has symmetry breaking points that are outside of the regime of the EFT which we conservatively set as $|\phi|>0.5\Lambda$. We extend the scan for Wilson coefficients up to $c_{1} = c_{2} = (4\pi)^{2}$. For values of $c_{12}$ larger than $O(1)$, we expect that the EFT series will not be convergent. With this truncation, however, we do not know what the maximum trusted value for $c_{12}$ will be. 

We can have a better understanding of the boundary between the s-confining region and symmetry breaking by using Sturm's theorem~\cite{10.5555/1197095} and obtaining the following no symmetry breaking condition:
\begin{align} \nonumber
&-29792 c_{12}^3 m_{12}^6-18 c_{12}^2 \kappa^2 m_{3/2}^4 (435 \gamma_{V} +226 \dot{\gamma}_{V}-336) \\
&+27 c_{12} \kappa^4 m_{3/2}^2 \left(81 \gamma_{V}^2+420 \gamma_{V}+32 \dot{\gamma}_{V}+248\right)+864 \kappa^6 > 0 \, .
\end{align}
For one-loop and small $\kappa$ we have the solution:
\begin{align}
c_{12} \approx \frac{3 \left(9+2 \sqrt{462}\right) \kappa^2}{266 m_{3/2}^2} \, .
\end{align}

We can see this peculiar effect that the symmetry breaking solutions for small couplings are moving outside of the EFT signaling that this region is still sensitive to higher-order terms and nothing can be said at this truncation level. Additionally, the existence of a small region with $\chi$SB inside the EFT regime looks promising, but at the same time, the values necessary of the Wilson coefficient could be a signal that we are outside of the convergence of the EFT and once again signaling the sensitivity of what happens at higher orders. As we vary $m_{3/2}$, the symmetry breaking solution maintains the same shape while undergoing rescaling wherein $\kappa$ transforms to $\kappa \, m_{3/2}/(0.1 \Lambda)$.  Consequently, we obtain the phase diagram corresponding to any value of $m_{3/2}$ from Figure~\ref{fig:SU2}. 

As in the $N_c=3$ case from Section~\ref{sec:Khaller3}, there is no clear way to perform matching of the Lagrangian parameters at $\Lambda$ to have an accurate better picture of what happens in the theory. Exploring higher-order terms could help better explore the phase diagram as we approach $\Lambda$, but it will quickly become intractable with the increase of the parameter space.
With higher order Kähler terms, we do expect a more complicated vacuum structure to emerge, particularly when terms like $\text{Pf}V\text{Pf}V^{\dagger}$ appear. This in turn becomes a problem~\cite{deLima:2022dht,Cho:1998vc} as we need to have more knowledge of the coefficients to be able to extract any useful information.

There could be some hope to be able to follow some specific Kähler terms from the UV to IR if they are related to conserved currents~\cite{Abel:2011wv} but this does not apply to a large class of operators. From our result, we can now see some regions of the parameter space with $\chi$SB that appear as a result of the inclusion of higher order Kähler corrections.

\section{Conclusion} \label{sec:conc}

In this work, we have presented a comprehensive study of the phase diagram of supersymmetric QCD-like theories with $N_{f}=N_{c}+1$ flavors perturbed by Anomaly Mediated Supersymmetry Breaking (AMSB). Our analysis extends previous studies of s-confining ASQCD theories in three different directions. Specifically, we have shown that the assumption that terms proportional to $m_{3/2}^{2}$ can be ignored when $m_{3/2}$ is small is not valid for the baryon preserving direction. Additionally, we have included two and three-loop contributions to the anomalous dimensions to investigate the robustness of our results and examined the leading effect of higher order Kähler corrections to investigate the stability of the phase diagram as we approach the confining energy scale.

We showed that the phase diagram of s-confining ASQCD is richer than one would naively estimate. The final answer including only the leading Kähler potential did not change, but for different reasons. Now the phase diagram shows the possibility of having a phase transition that restores chiral symmetry. We show that the theory never reaches this state and at the pertubativity boundary the theory is in the symmetry breaking phase. This applies for all s-confining ASQCD with $N_{c}>2$.

The QCD-like vacuum depends on the non-renormalizable coupling in a peculiar way. It moves the solution away from the EFT regime in the IR. The dynamic is different when this coupling is zero where the theory goes to the symmetric phase. This signals that we cannot consider only the $\widetilde{B} M B$ operator in the deep IR. The dynamic never reduces to the classically conformal limit. It is still the coefficient of the $\widetilde{B} M B$ operator that defines the phase of the theory, but the coefficient of the $\det M$ operator is the one that defines if the solution is inside or outside the EFT regime where we have theoretical control. 
 
We included the next-leading-order Kähler corrections for $N_{c}=3$. We see that as we include the higher-order terms there is the possibility of a broken baryon vacuum. Nevertheless, we expect the theory to be close to the perturbativity boundary as we approach $m_{3/2} \approx \Lambda$, and thus we expect that the theory remains in the QCD-like  $\chi$SB phase. 

It is important to emphasize that currently, we cannot definitively prove that nothing extraordinary will happen with the theory as we move into the strong SUSY breaking regime with $m_{3/2} \sim \Lambda$. However, the theories with finite $m_{3/2}$ remain well-defined confining theories. We now have a better understanding of the anticipated spectrum of these theories. Further investigation is required for extrapolating to large $m_{3/2}$, but it is intriguing that the spectrum of the theory for small $m_{3/2}$ aligns with our expectations for the theory with large $m_{3/2}$.

In Section~\ref{sec:Khaller2}, we focused on the special case of $N_{c}=2$. The spectrum derived from small SUSY breaking does not match the expectation that the non-supersymmetric version of the theory will have $\chi$SB. We review this result and extend the analysis to include higher order loop corrections and the next order Kähler corrections. The hope is that it is possible to obtain information on the phase structure of the theory inside the region of validity of AMSB, even if the transition occurs at large SUSY breaking. 

 The phase diagram for $SU(2)$ s-confining ASQCD changes slightly with the inclusion of NLO Kähler terms, now having a small region that has chiral symmetry breaking.  These solutions were mostly outside of the regime of the EFT, signalling the dependence on the dynamics around $\Lambda$. Regions with $\chi$SB with higher Yukawa values were inside the EFT domain but had relatively large Wilson coefficients, where the EFT convergence can be problematic.  Because of this, it is not possible to draw any conclusions on what happens with this theory without a deeper understanding of the Kähler coefficients.  

It is important to point out that the $SU(2)$ s-confining AMSB theory raises an important concern on the applicability of the Anomaly Mediated Supersymmetry Breaking (AMSB) method in investigating confining theories. While the AMSB theory provides a profound understanding of the low-energy and UV degrees of freedom of a class of nearly supersymmetric theories, its accuracy in replicating QCD results is contingent on its ability to increase the SUSY breaking beyond the confining scale to fully integrate out the SUSY degrees of freedom. Without the capability to identify the correct symmetry breaking pattern based on the behavior of small $m_{3/2}$, the efficacy of AMSB as a tool for understanding non-SUSY theories becomes limited.  
 
Further exploration into these kinds of theories where there should be symmetry breaking and AMSB hints at moduli stabilized at the origin
is necessary to understand if we can systematically use ASMB as a tool to study broad classes of strongly coupled field theories. In this work, we perform the first step in this direction, but there is much more to explore.

\section*{Acknowledgments}
\noindent We would like to thank Yang Bai, Benjamin Grinstein, Julio Parra-Martinez, Yael Shadmi, Yuri Shirman, and Ofri Telem for useful discussions. C.H.dL.~and D.S.~are supported in part by the Natural Sciences and Engineering Research Council of Canada (NSERC). C.H.dL is supported in part by the 2023 IPP Early Career Theory Fellowship.

\appendix

\section{Loop Calculations} \label{app:not}

In this paper, we use the conventions for anomalous dimension following \cite{Murayama:2021xfj}. It is important to note that these conventions are different from those used in~\cite{Martin:1997ns,Martin:1993zk,Jack:1996qq}, which is where we obtained the two and three-loop results. We can convert between these conventions by multiplying the results of ~\cite{Martin:1997ns,Martin:1993zk,Jack:1996qq} by $(-2)$. For a theory with a superpotential given by $W = \frac{1}{3!}y_{ijk}\phi_{i}\phi_{j}\phi_{k}$, we can write the anomalous dimensions in our conventions as:
\begin{align}
\gamma_{i}^{(1)} &= -\frac{1}{16\pi^{2}} \left( y^{ijk}y_{ijk} \right)  \, , \\ 
\gamma_{i}^{(2)} &= \frac{1}{(16\pi^{2})^{2}} \left( y_{imn}y^{mpi}y^{nkl}y_{pkl} \right) \, , \\
\gamma_{i}^{(3)} &= -\frac{1}{(16\pi^{2})^{3}} \left(3\zeta(3)M^{i}_{i} + 4 \left(y.S_{4}.y\right)^{i}_{i} - S_{7 \, i}^{i}-2S_{8 \, i}^{i} \right) \, ,
\end{align}
where the index $i$ is fixed and we are using the following definitions:
\begin{align}
M^{i}_{j} &= y^{ikl}y_{kmn}y_{lrs}y^{pmr}y^{qns}y_{jpq} \, , \\
S^{i}_{4 \, j}&= \frac{1}{2}y^{imn}y^{pkl}y_{mkl}y_{jpn} \, , \\ 
S^{i}_{7 \, j} &= \frac{1}{4} y^{imn} y^{pkl}y_{mkl}y^{qst}y_{nst}y_{jpq} \, , \\
S^{i}_{8  \, j} &= \frac{1}{4} y^{imn} y^{pkl}y_{qkl}y^{qrs}y_{mrs}y_{jpn} \, .
\end{align}

\section{Second minimum of $N_{c}>2$ s-confining ASQCD} \label{app:secondMIN}

\begin{figure*}[h!]
   \resizebox{0.45\linewidth}{!}{\includegraphics{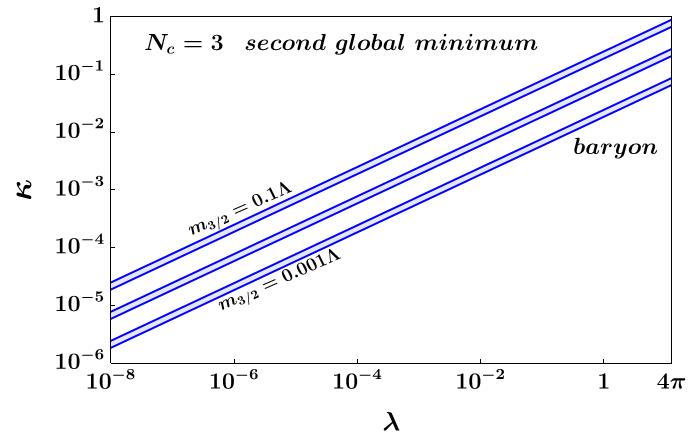}}
      \resizebox{0.45\linewidth}{!}{\includegraphics{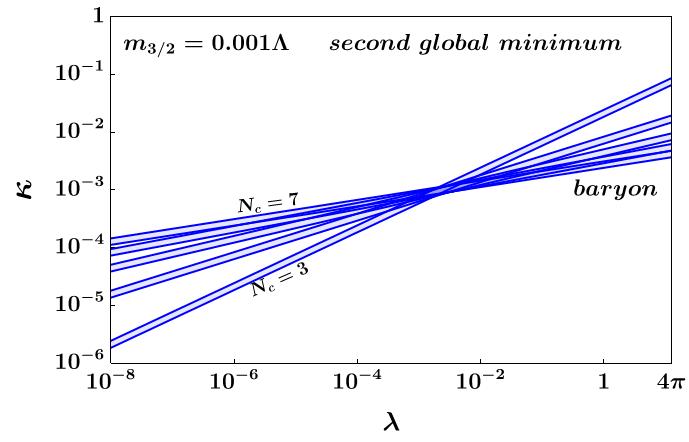}}
     \caption{ \label{fig:NUMSECOND}Phase diagram at three-loops for the second lowest minimum. On the left side we vary $m_{3/2}$ and fix $N_{c} = 3$, and the right panel fixes $m_{3/2}=0.001\Lambda$ while varying $N_{c}$. The small blue regions show the region where the second minimum is baryon breaking, while the rest of the region has a second minimum at the origin.}
\end{figure*}
As noted in Section~\ref{sec:phase}, there can be a local minimum in the $N_c>2$ theory that breaks baryon number, $b\neq 0$, even with only a canonical Kähler potential. In our scans, we noticed that this minimum, if it exists, is always above one of the two main minima discussed in Section~\ref{sec:phase}. We want to highlight here some peculiar features of these baryon breaking local vaccua which were only possible to explore numerically.  

This region of parameter space is better explored on a log scale. This baryon breaking vacuum only appears in a small band in the $\lambda-\kappa$ plane for fixed $m_{3/2}$ and $N_c$.
We performed a scan using a log-uniform random distribution looking only for the second minimum which can be seen in Figure~\ref{fig:NUMSECOND}. 
One important result that we obtained from our scans is that we never had any two minimums of the same type.  This means that, at most, we obtained three configurations existing at the same time, s-confining, broken baryon, and the QCD-like $\chi$SB vacuum. The region where the local broken baryon vacuum exists exhibits a characteristic power-law scaling of $\kappa \propto \lambda^{1/(N_{c}-1)}$~\footnote{This scaling only works for small $m_{3/2}$ and starts to deviate when we go above $m_{3/2}=0.1\Lambda$. } and has non-trivial dependence on $m_{3/2}$. It is interesting to note that this solution does not have the same peculiar scaling of $1/\lambda$ of the baryon preserving direction in Eq.~\eqref{eq:MIN}. This means that, for sufficiently small $\lambda$, the QCD-like $\chi$SB vacuum is outside the region with theoretical control, leaving only this baryon breaking minimum inside. 

\begin{figure*}[t!]
    \resizebox{0.32\linewidth}{!}{\includegraphics{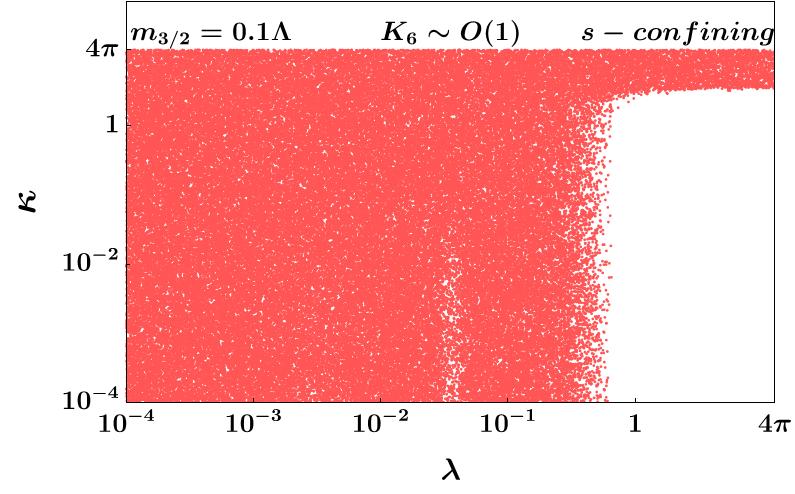}}
    \resizebox{0.32\linewidth}{!}{\includegraphics{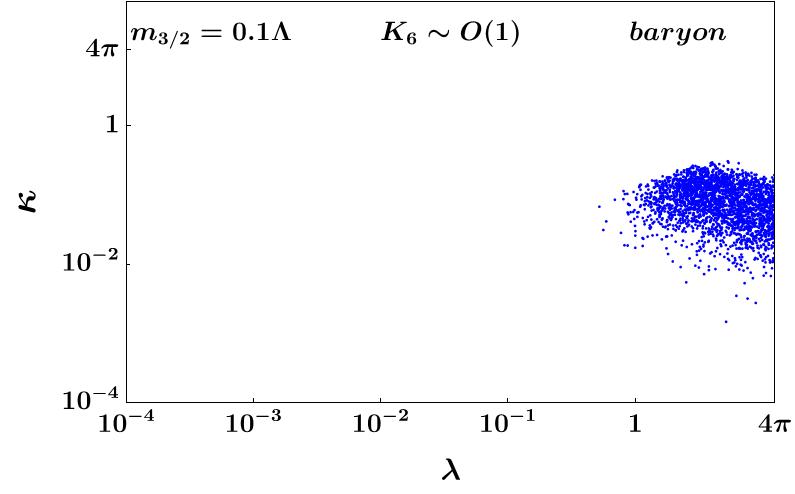}}
    \resizebox{0.32\linewidth}{!}{\includegraphics{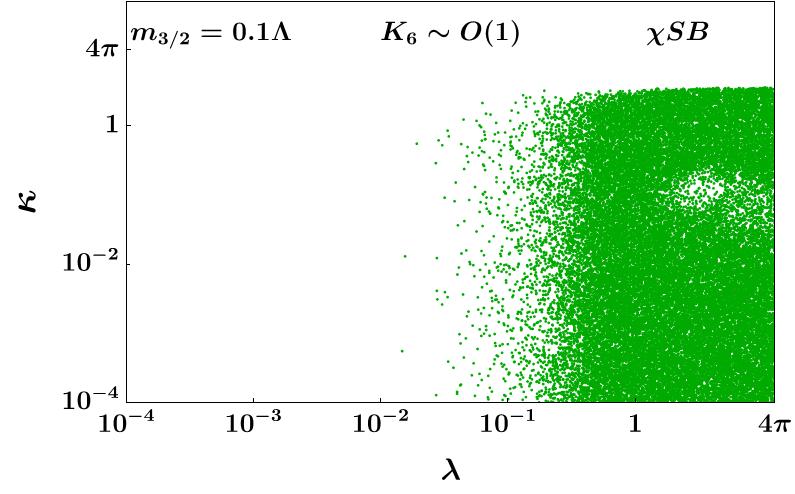}}
     \caption{ \label{fig:SCANLOG} Same as Figure~\ref{fig:KHASCAN}, but with $\kappa$ and $\lambda$ on a log scale.
     }
\end{figure*}

We also explored if broken baryon solutions from the Kähler corrections introduced in Section~\ref{sec:Khaller3} are deformations of this solution from Figure~\ref{fig:NUMSECOND}. We can see in Figure~\ref{fig:SCANLOG}, which is the same as Figure~\ref{fig:KHASCAN} but on a log scale for $\kappa$ and $\lambda$, that this is not the case. While we can see that there is a small region where there is no QCD-like $\chi$SB and the only global minimum is the broken baryon solution.

\bibliographystyle{apsrev-title}
\bibliography{bibSUSY}

\end{document}